\documentclass[aps,twocolumn]{revtex4-1}

\usepackage[utf8]{inputenc}
\usepackage{amsmath}
\usepackage{stmaryrd}
\usepackage{color}
\usepackage{mathrsfs}
\usepackage{yfonts}
\usepackage{rotating}
\usepackage{graphicx}
\usepackage{amsfonts}
\usepackage{mathrsfs}
\usepackage{amsmath}
\usepackage[american]{babel}
\usepackage{wasysym}
\usepackage{slashed}
\usepackage{pgfplots}
\usepackage{braket}
\usepackage{caption}
\usepackage{graphicx}
\usepackage{subfigure}
\usepackage{dsfont}

\newcommand{\g}[0]{\gamma}
\newcommand{\al}[0]{\alpha}
\newcommand{\be}[0]{\beta}

\newcommand{\rrho}[0]{\hat{\rho}}

\newcommand{\ma}[1]{\mathcal{#1}}

\newcommand{\ea}[1]{\begin{align}#1\end{align}}
\newcommand{\eq}[1]{\begin{equation}#1\end{equation}}

\begin{document}
\title{Classifying Quantum Entanglement through Topological Links}
\author{Gon\c{c}alo M. Quinta}
\email{goncalo.quinta@tecnico.ulisboa.pt}
\affiliation{Centro de Astrof\'{\i}sica e Gravita\c c\~ao  - CENTRA,
Departamento de F\'{\i}sica, Instituto Superior T\'ecnico - IST,
Universidade de Lisboa - UL, Av. Rovisco Pais 1, 1049-001 Lisboa, Portugal}
\author{Rui Andr\'{e}}
\email{rui.andre@tecnico.ulisboa.pt}
\affiliation{Centro de Astrof\'{\i}sica e Gravita\c c\~ao  - CENTRA,
Departamento de F\'{\i}sica, Instituto Superior T\'ecnico - IST,
Universidade de Lisboa - UL, Av. Rovisco Pais 1, 1049-001 Lisboa, Portugal}


\begin{abstract}
We propose a new classification scheme for quantum entanglement based on topological links.
This is done by identifying a non-rigid ring to a particle, attributing the act of cutting and removing a ring to the operation of tracing out the particle, and associating linked rings to entangled particles.
This analogy naturally leads us to a classification of multipartite quantum entanglement based on all possible distinct links for a given number of rings.
To determine all different possibilities, we develop a formalism which associates any link to a polynomial, with each polynomial thereby defining a distinct equivalence class.
In order to demonstrate the use of this classification scheme, we choose qubit quantum states as our example of physical system. A possible procedure to obtain qubit states from the polynomials is also introduced, providing an example state for each link class.
We apply the formalism for the quantum systems of three and four qubits, and demonstrate the potential of these new tools in a context of qubit networks.
\end{abstract}
\maketitle
\vspace{2mm}

\section{Introduction}

The prospect of quantum technologies is deeply dependent on our understanding of entanglement. Applications like quantum teleportation, quantum computation and quantum cryptography are built upon the properties of entanglement. However, although these ideas have been successfully tested in laboratory, the hope of fully harnessing their potential is strongly tied to our understanding of multipartite entanglement.



One of the most difficult aspects of quantum entanglement classification is connected to the multitude of criteria one may use to define different classes. A particular natural choice of equivalence class is the group of reversible stochastic local quantum operations assisted by classical communication (SLOCC) operations, which essentially views two entangled states as equivalent if one of them can be obtained from the other with some finite probability, by use of local measurements alone. Both these states can then be used, in principle, for the same quantum information processing tasks, albeit with different probabilities of success. This idea has been successfully applied to the characterization of three-qubit states in \cite{Acin:2000, Dur:2000}. Classification of four-qubit states has received some attention \cite{Verstraete:2002, Lamata:2007, Li:2009, Borsten:2010} and methods for the determination for more general number of qubits have also been developed \cite{Viehmann:2011, Sharma:2012}. Depending on the criteria used for classification, the total number of four-qubit entanglement classes range from 6~\cite{Sharma:2012} to 49~\cite{Li:2009}.

The purpose of this work is to study quantum entanglement in a different way, by formalizing the tentative analogy between quantum entanglement and topological links. The similarities between the two have already been noticed by other authors in the past \cite{Aravind:1997, Sugita:2007, Kauffman:2002}, although it has not been developed much further. Aravind \cite{Aravind:1997} was the first to publish the remark, pointing out the similarities between three particle quantum states and different ways to link three rings. He suggested that if each particle was associated to a ring, then there would be three different ways to link all of the three rings, corresponding to three different entanglement classes. The act of measuring a particle state could then be equivalent to cutting the corresponding ring, and if the remaining rings were still linked it would indicate the corresponding remaining particles were still entangled. The latter reasoning was limited, however, as he noted that performing the measurement in different basis would not lead to the same conclusions. This limit in the analogy was dealt with by Sugita \cite{Sugita:2007}, where he suggested the act of cutting the ring to be associated with the basis independent operation of tracing out the correspondent particle from the density operator, which physically corresponds to viewing the system as if the particle did not exist. Furthermore, the use of the trace operation facilitates the incorporation of quantum systems with more then two levels, which were the only ones considered until then.

In this paper we use topological links as a tool to study quantum entanglement. By associating each particle to a non-rigid ring and using as classification criteria the different ways to link all rings, we arrive at a set of entanglement classes. The new element which allows us to develop a mature connection between links and quantum entanglement is the way we define equivalence between links, which focuses not on the way each ring is linked but rather on whether a ring is linked or not to each other ring. The appeal of this classification resides in the intuitive picture of quantum entanglement it provides, predicting a large amount of equivalence classes as the number of particles grow. We choose to incorporate these ideas using qubits as the quantum systems, for their simplicity and technological applications. On a physical perspective, we show that the immediate applications this new classification scheme finds in quantum information are not related to direct protocol applications, like SLOOC classification, but rather to control which parties within a quantum network will not be able to successfully execute protocols between each other.

This paper is organized as follows. In Sec.~\ref{form} we develop a set of rules which attributes a polynomial to each link class, thereby making it possible to find out how many distinct links there are. In Sec.~\ref{quantcla} we establish a procedure which identifies which link corresponds to a given state, as well as a way to determine a qubit state associated to a given link. In Sec.~\ref{3q} we apply the previous tools in detail for three qubits, while Sec.~\ref{4q} will be dedicated to the four qubit case. In Sec.~\ref{linkrep} we provide a simple method to depict a link by using its associated polynomial. In Sec.~\ref{physapp} we highlight an example where entanglement classification using links provides new applications in quantum information, namely on qubit networks. Finally, in Sec.~\ref{conc} we draw the conclusions.

\section{A classification of topological links}
\label{form}


In this section we develop the formalism used to address the following question: in how many ways can a given number of rings be linked? First of all, this question must be made rigorous from a knot theory perspective. The term ``ring'' can be interpreted as a knot, so that linked rings can be seen as linked knots. We shall denote $n$ linked rings simply as $n$-ring links. In fact, there are infinite ways to link even two rings, but if we address the problem by finding which rings remain linked after any given ring is cut, then any two $n$-ring links can be considered equivalent if the results of all possible combinations of cuts are the same.

To put the above idea to practice, a particular configuration of any number of linked rings will be characterized by a polynomial, hereby denoted as $\ma{P}$.
The construction of this polynomial starts by associating a variable to each
ring and then interpreting the product between variables to be equivalent to the associated rings being linked. Taking a variable to 0 is then interpreted as the associated ring being cut. A given link can thus be characterized by the remaining links produced from all possible cuts.

\begin{figure}[h!]
  \centering
    \includegraphics[width=0.20\textwidth]{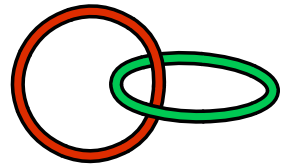}
      \caption{The Hopf link, an element of the class $2^1$.}
      \label{21}
\end{figure}

Let us start with the simplest example, consisting of two rings. There is only one link class associated to them, which we designate $2^1$. The notation $n^i$ indicates that we are considering the $i^{\rm th}$ link class with $n$ rings, where the value of $i$ is a label to identify which link we are referring to (all the labels for three and four rings will be explicit in the next sections). The Hopf link depicted in Fig.~\ref{21}, for example, is an element of this class. We used the KnotPlot software to generate all representations of links in this work. When representing a link, we shall adopt the color red for the variable $a$, green for $b$, blue for $c$ and yellow for $d$.
The configuration of Fig.~\ref{21} could be represented in any other way, as long as the two rings were linked; what matters is if the rings are linked or not. To summarize this information in a concise way, we associate to this link the polynomial
\eq{
\ma{P}({2^1}) = ab\,.
}
Henceforth designated ``ring variables'', the letters $a$ and $b$ designate each of the rings, and the product of these variables will be interpreted as the two rings being linked. The ring variables do not need to belong to any particular space, so we will take the latter to be the set of real numbers. Now, cutting a ring will be taken as setting the correspondent variable to 0, so cutting any of the rings from a $2^1$ link will always result in the polynomial 0, which shall thus represent the case were no rings are linked. In case we want to consider multiple disconnected links, we use the direct sum symbol, such that $2^1 \oplus 2^1$, for example, would correspond to two disconnected $2^1$ links.

\begin{figure}[h!]
  \centering
    \includegraphics[width=0.30\textwidth]{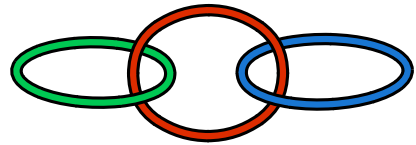}
      \caption{An element of the link class $3^3$.}
      \label{33}
\end{figure}

As a more complex example, consider three rings, denoted by $a$, $b$ and $c$, and the polynomial $ab + ac$. The problem now is to identify the corresponding link. In this case, the ring $b$ is linked to ring $a$ and the ring $c$ is also linked to $a$, while the rings $b$ and $c$ are not directly linked. Now, cutting the ring $b$, i.e. taking $b=0$, we are left with $ac$, meaning the two rings $a$ and $c$ are still linked.
The same thing happens when the ring $c$ is cut, which leaves the term $ab$, implying that the rings $a$ and $b$ remain linked. All the previous remarks are enough to generate a visual representation of the link, which resembles a chain, depicted in Fig.~\ref{33}. We denote this link class by $3^3$.

\begin{figure}[h!]
  \centering
    \includegraphics[width=0.11\textwidth]{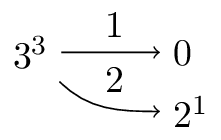}
      \caption{Diagram of 1-cuts for the link class $3^3$. The arrows correspond to the link after the cut, and the numbers above each arrow represent the number of possible cuts which lead to a link of that class.}
      \label{d33}
\end{figure}

The information associated to the ring cuts will also be described using a diagram, exemplified in Fig.~\ref{d33}, providing a quick summary of the link's properties. A diagram representing $n$ cuts will also be denoted as a $n$-cut diagram. For notational convenience, it does not give detailed information as to which rings remain linked, so this does not fully characterize a link since two different links may have the same diagram.

Intuitively, there are a number of rules that should be satisfied in order for a certain polynomial to represent a valid link. For example, terms with repeated variables, like $aab$, should not appear, as it represents superfluous information; we already know that the ring $a$ is linked to $b$ from the product $ab$, so multiplying another $a$ is redundant. Consequently, one only needs to use terms where no variables are repeated.  In addition, in this work we will only be concerned with the case where all initial rings are linked, so all ring variables should appear at least once. This also means the possibility of including single variable terms is not of interest to us, since it would mean that the ring is not linked to any other. For instance, for three rings the set of all possible terms is $\{abc,ab,ac,bc\}$, so any polynomial can be written as a linear combination of the elements of this basis. For simplicity, we shall refer to an $n$-variable term as $n^{\rm th}$ order term. For $N$ rings, one has $\binom{N}{2}$ second order terms, $\binom{N}{3}$ third order terms and so on, up to $\binom{N}{N}=1$ $N^{\rm th}$ order terms, adding to a total of $2^N-N-1$ terms. Furthermore, no particular importance is given to any ring variable since no information is contained in the variables themselves. For example, the polynomials $ab+ac$ and $ab+bc$ represent the same link class. Finally, there is one last rule, related to the submonomials of a given term. Considering the case where a given polynomial contains the terms $ab$ and $ac$, it is irrelevant if we further add the term $abc$, since all the letters of the latter term have already appeared in the smaller monomials. Indeed, one may check by setting each letter to zero independently, that the results of all possible cuts are equal for both $ab+bc$ and $ab+bc+abc$, and so both polynomials must represent the same link class. However, consider we have yet another ring $d$ and construct the links $ab+cd$ and $ab+cd+abcd$. Clearly they have different properties, in the sense that the first one corresponds to two disconnected 2-ring links, while the former represents a 4-ring link. The fact that the results of each cut are different for the two polynomials also emphasizes this. It becomes clear that the fourth order monomial is only discardable if the second order monomials constitute a 4-ring link.

We summarize all of the above remarks in the following set of rules for the polynomials:
\begin{itemize}
\item[1)] There must not be any repeated terms, i.e. no ring variable can have a power greater than 1;
\item[2)] Each ring variable must appear at least once;
\item[3)] There must not be first order terms;
\item[4)] Relabeling of variables is irrelevant;
\item[5)] An $n$-variable monomial $M$ is irrelevant if all of its variables are already present as an $n$-ring link of lesser order monomials, built only with the variables of $M$.
\end{itemize}
These rules are sufficient to select the desired polynomials, corresponding to each distinct link class, for any number of rings. By construction, any link can be associated to one and only one of these classes.

\section{Quantum entanglement classification through topological links}
\label{quantcla}

In this section we shall firstly demonstrate how one can attribute a link class to any given quantum state, providing an intuitive picture of the entanglement properties of the state under the trace of particles.
After this, we will show how to find a qubit quantum state associated to a given link class, a task which is particularly important when one is primarily motivated by the properties of a given link.

\subsection{From quantum states to links}
\label{qutolink}

We now turn to the classification procedure which associates a link to a given quantum state. In order to find it, one needs to perform all possible qubit traces on the state and check, for each result, if the traced out state is entangled or not. This is analogous to performing all possible ring cuts for a given link and see if the remaining rings remain linked or not. As a result, the collected information from the traces can be used to identify the link class.

As an example, consider the 3-qubit state in the computational basis
\eq{\label{33ket}
\ket{\psi} = {1\over 2}\left(\ket{1 0 0}_{abc}+\ket{0 1 0}_{abc}+\ket{1 1 0}_{abc}+\ket{0 1 1}_{abc}\right)\,,
}
with density operator $\rrho = \ket{\psi}\bra{\psi}$. We associate the qubit $a$ to the ring identified by the ring variable $a$, and so on. First, since we will solely be interested in the case where all qubits are entangled in the initial state, we must check for tripartite entanglement. This can be done by performing a positive partial transpose (PPT) test on each pair of possible subsystems, i.e. we must partially transpose individually the qubits $a$, $b$, and $c$ and investigate the sign of the corresponding eigenvalues of each operator. We will denote $\rrho^{T_a}$ as the partial transposition of $\rrho$ with respect to the subsystem $a$. Negativity of at least one eigenvalue will ensure that the two subsystems are entangled \cite{Peres:1996}. Non-negativity will not guarantee that the subsystems are separable since PPT tests for separability are only sufficient for $2\times 2$ or $3\times 2$ systems. In this regard, one simply changes the initial state until we are in a situation where the PPT test guarantees a tripartite entangled state. For the case of $\ket{\psi}$, one may check that all subsystems are entangled, and so the state has tripartite entanglement, i.e. all three rings are linked. Next, we trace out at a time the qubits $a$, $b$ and $c$, giving the two qubit density operators $\rrho_{bc}$, $\rrho_{ac}$ and $\rrho_{ab}$. Performing again a PPT test on each of these operators will be sufficient to determine if the subsystems are entangled or separable, since they are $2\times 2$ systems. We find that only $\rrho^{T_a}_{ab}$ and $\rrho^{T_a}_{ac}$ have negative eigenvalues, meaning that when the rings $b$ or $c$ are cut, the remaining rings cannot be pulled apart, while the opposite happens when the ring $a$ is cut. This is exactly the behavior depicted by the link of Fig.~\ref{33}, so we say the state is of the type $3^3$.

The described procedure becomes more complicated for more than two qubits, for two reasons: the number of possibilities after each cut increases; and the lack of a sufficient criterion for checking separability of mixed states in systems other than $2\times 2$ or $2\times 3$. Both of these topics will be addressed in Sec.~\ref{4q}.

\subsection{From links to quantum states}
\label{qumap}

While we may find a link to each quantum state, the opposite task is also important. This becomes clearer as the number of qubits increases. For example, it would not be a simple task to obtain a four qubit state with entanglement properties described by the polynomial $abcd+abc+ab$, without any reduction of the initial degrees of freedom.  It would thus be very useful to have a map enabling us to immediately write a state with such properties directly from the polynomial. Although a full map is not yet available, we develop in this section an example of an incomplete map, in the sense that the basic structure of the state can be written down but some computational power is still needed to fix the remaining free coefficients. This is already a large step, as it greatly reduces the initial free coefficients of the initial general state, which grows as $2^N$, where $N$ is the number of qubits. Ideally, it would be optimal to have a method which associated a pure state for each link. However, although we were able to find pure states for all 3-ring links, the same has not been achieved yet for the 4-ring case. As such, we shall demonstrate a procedure which provides mixed states only.

The idea is to attribute to each term in the polynomial a mixed state in the computational basis, with the full state being the sum of all the individual states for each term of the polynomial. In this paper we will work up to four qubits and will use as building blocks the GHZ type states~\cite{GHZ:1989}
\ea{
\ket{2^1}_{ij} & = {1 \over \sqrt{2}} \left(\ket{0 0}_{ij} + \ket{1 1}_{ij}\right)\,, \\
\ket{3^1}_{ijk} & = {1 \over \sqrt{2}} \left(\ket{0 0 0}_{ijk} + \ket{1 1 1}_{ijk}\right)\,, \\
\ket{4^1}_{ijkl} & = {1 \over \sqrt{2}} \left(\ket{0 0 0 0}_{ijkl} + \ket{1 1 1 1}_{ijkl}\right)\,,
}
or, in general,
\eq{\label{Nket}
\ket{N^1}_{ij\ldots} = {1 \over \sqrt{2}} \left(\ket{0 0 \ldots}_{ij\ldots} + \ket{1 1 \ldots}_{ij\dots}\right)\,,
}
where $N$ denotes the number of qubits. We start by selecting any term of the polynomial and associate to it a state of the form
\eq{\label{GenStru}
\ket{{\rm Entangled \,\, qubits}}\ket{{\rm Separable \,\, qubits}}\ket{{\rm Extra \,\, qudit}}\,,
}
where $\ket{{\rm Entangled \,\, qubits}}$ is an entangled state of the form of Eq.~(\ref{Nket}) made of the qubits associated to the ring variables appearing in the term, $\ket{{\rm Separable \,\, qubits}}$ is a separable state containing the qubits associated to the ring variables which do not appear in the term and $\ket{{\rm Extra \,\, qudit}}$ is a qudit state introduced for notational convenience, whose purpose is to be traced out in the end. For instance, consider again the polynomial $ab + ac$ for a 3-ring link. Using the structure of Eq.~(\ref{GenStru}), the state associated to the term $ab$ can be written as $\ket{2^1}_{ab}\ket{q_1}_c\ket{0}_d$ where the value of $q_1$ is not specified and the qudit subscript is always the next letter in alphabetic order after the last qubit. As for the term $ac$, the associated state is $\ket{2^1}_{ac}\ket{q_2}_{b}\ket{1}_d$, where the qudit is increased by one unit. The structure of the full mixed state associated to the polynomial $ab + ac$ can thus be written as
\eq{\label{rhopsi}
\rrho_{abc}(3^i) = {\textrm{tr}_d \left[\ket{\psi_i}\bra{\psi_i}\right] \over \sqrt{\braket{\psi_i | \psi_i}}}\,,
}
where
\eq{\label{33state}
\ket{\psi_i} = c_1 \ket{2^1}_{ab}\ket{q_1}_c\ket{0}_d + c_2 \ket{2^1}_{ac}\ket{q_2}_{b}\ket{1}_d\,.
}
With the notation of Eq.~(\ref{rhopsi}), in order to specify a state we will only need to write down the state $\ket{\psi_i}$. At this point, one must resort to computational methods in order to determine the coefficients $c_1$ and $c_2$ and the value of the free qubits $q_1$ and $q_2$. Nonetheless, we shall find that for three and four qubits the coefficients can usually be set to $1$, with only a few exceptions where they take the value $1/2$. Regarding the separable qubits, there is usually a large number of possibilities, so it does not take a lot of computational power to find one possible state. We will discuss these topics in more detail in Secs.~\ref{3q} and \ref{4q}.

Finally, as previously remarked, the analogy between links and quantum entanglement does not restrict itself to qubits. In fact, we could have considered any other physical observable $A$ with a general spectrum $\{\al, \be, \g, \ldots\}$ and three particles labeled by $a$, $b$ and $c$. One could then define, for example, the state
\eq{
\ket{\phi} = \ket{\be \al \g}_{abc} + \ket{\be \al \be}_{abc} + \ket{\g \g \al}_{abc}\,,
}
and perform all possible partial traces in order to identify the 3-ring link class associated to the quantum state. This would also be true even if the observable $A$ had a continuous spectrum, where the only difference would be in the definition of the trace operation. The choices of identifying a particle with a ring and a cut with the partial trace of a particle are basis independent, so the connection between links and quantum entanglement is independent of the physical system being considered. As a consequence, the same way that there is a 3-ring link identifying the three qubit mixed state of Eq.~(\ref{rhopsi}), there will also be a 4-ring link associated to the pure state in Eq.~(\ref{33state}) with four qubits.

\section{Application to three qubits}
\label{3q}

In this section we shall be concerned with the different classes of entanglement for 3 qubits, regarding only the case where all qubits are entangled in the initial state. Following the premise of this paper, this is equivalent to finding the number of ways in which three rings can be linked, which in turn corresponds to finding the different types of polynomials one may construct from ring variables. We shall see that the results obtained in this section not only contain those of previous works \cite{Aravind:1997, Dur:2000}, but also a new one which, as far as the authors are aware, has not yet been documented in the literature. The 3-qubit case is also the only one where both a pure and a mixed state will be provided for all classes.

We start by constructing the basis of possible terms to be used, which corresponds to $\{abc, ab, ac, bc\}$. The following 4 distinct types of polynomials
\ea{
\ma{P}({3^1}) & = abc\,, \\
\ma{P}({3^2}) & = abc + ab\,, \\
\ma{P}({3^3}) & = ab + ac\,, \\
\ma{P}({3^4}) & = ab + ac + bc \label{W}\,,
}
can be obtained either by direct inspection or by adapting the rules of polynomial construction of Sec.~\ref{form} to some symbolic manipulation program. Let us now analyze individually each of the link classes.

\begin{figure}[h!]
  \centering
    \includegraphics[width=0.11\textwidth]{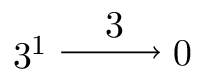}
      \caption{The 1-cut diagram of the link $3^1$.}
      \label{d31}
\end{figure}

\begin{figure}[h!]
  \centering
    \includegraphics[width=0.25\textwidth]{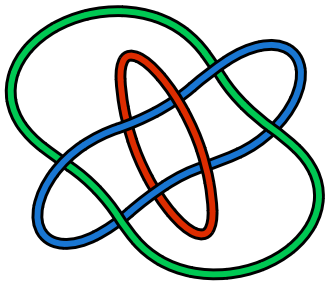}
      \caption{The Borromean link, an element of the link class $3^1$.}
      \label{31}
\end{figure}

Starting with the link $3^1$, taking any variable to 0 will result in 0, i.e. cutting any of the rings will result in setting all the remaining rings free. The respective 1-cut diagram is represented in Fig.~\ref{d31}.
This is the characteristic quality of the well known Borromean link, shown in Fig.~\ref{31}. The generalization for any number of rings is known as the Brunnian link.
Using the procedure of Sec.~\ref{qumap} is not strictly necessary for this case, as this type of behavior is well known to be associated to the GHZ type states of Eq.~(\ref{Nket}), for any number of qubits. Hence, we have the pure state
\eq{
\ket{3^1}_{abc} = {1 \over \sqrt{2}} \left(\ket{0 0 0}_{abc} + \ket{1 1 1}_{abc}\right)
}
or, for completeness, the state constructed by inserting in Eq.~(\ref{rhopsi}) the state
\eq{
\ket{\psi_1} = \ket{3^1}_{abc}\ket{0}_{d}\,.
}

\begin{figure}[h!]
  \centering
    \includegraphics[width=0.11\textwidth]{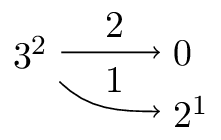}
      \caption{The 1-cut diagram of the link class $3^2$.}
      \label{d32}
\end{figure}

\begin{figure}[h!]
  \centering
    \includegraphics[width=0.25\textwidth]{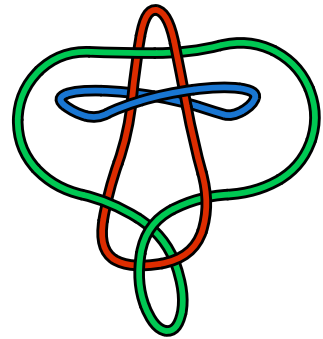}
      \caption{An element of the link class $3^2$.}
      \label{32}
\end{figure}

Moving on to the link class $3^2$, it exhibits the following behavior: when the rings $a$ or $b$ are cut, all other rings are set free, while if $c$ is cut, $a$ and $b$ remain linked. This information is summarized in Fig.~\ref{d32} and visually represented in Fig.~\ref{32}.
An example of a pure state is found to be
\eq{
\ket{3^2}_{abc} = {1\over \sqrt{3}}\left(\ket{000}_{abc} + \ket{111}_{abc} + \ket{001}_{abc}\right)\,,
}
while a mixed state can be constructed from
\ea{
\ket{\psi_2} = \ket{3^1}_{abc}\ket{0}_d + \ket{2^1}_{ab}\ket{0}_c\ket{1}_d\,.
}
Unlike all other three classes appearing in this section, this class has never been mentioned in the literature. This may be mostly because the approach taken in \cite{Aravind:1997, Sugita:2007} involved directly picturing the links, a process which quickly becomes non-trivial. Using the polynomial approach, however, it arises naturally.

The link class $3^3$ has already been treated in the previous section, depicted by Fig.~\ref{33}, and an associated pure state has already been given in Eq.~(\ref{33ket}). A mixed state can be constructed from Eq.~(\ref{33state}), more specifically as
\eq{
\ket{\psi_3} = \ket{2^1}_{ab}\ket{0}_c\ket{0}_d + \ket{2^1}_{ac}\ket{1}_b\ket{1}_d\,.
}
This type of state has also been suggested in \cite{Aravind:1997} to be associated to the link configuration resembling a chain.

\begin{figure}[h!]
  \centering
    \includegraphics[width=0.11\textwidth]{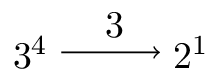}
      \caption{The 1-cut diagram of the link class $3^4$.}
      \label{d34}
\end{figure}

\begin{figure}[h!]
  \centering
    \includegraphics[width=0.25\textwidth]{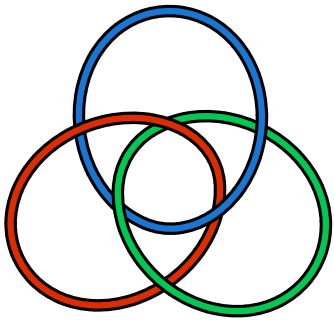}
      \caption{An element of the link class $3^4$.}
      \label{34}
\end{figure}

Finally, the link class $3^4$ has somewhat the inverse properties of the Borromean link $3^1$, i.e. setting any variable to 0 will result in a single two variable term. Equivalently, cutting any ring will always leave two linked rings. This leads to the 1-cut diagram in Fig.~\ref{d34} and an element of the class is depicted in Fig.~\ref{34}.
A pure state associated to this link class has been studied before and is know as the W state \cite{Dur:2000}, given by
\eq{
\ket{3^4}_{abc} = {1 \over \sqrt{3}} \left(\ket{001}_{abc} + \ket{010}_{abc} + \ket{100}_{abc}\right)\,.
}
The properties of this state's structure have also been connected before to the same link in \cite{Aravind:1997}. A mixed version of it can be created by using the state
\eq{
\ket{\psi_4} = \ket{2^1}_{ab}\ket{0}_c\ket{0}_d + \ket{2^1}_{ac}\ket{1}_b\ket{1}_d + \ket{2^1}_{bc}\ket{0}_a\ket{2}_d\,.
}
It can be shown that the density operators resulting from all states verify the expected behavior for all individual qubit traces, as required. This task is straightforwardly implemented in any algebraic manipulation software and is particularly simple to verify. 

It is also interesting to note that, from the point of view of cut diagrams, the classes $3^1$ and $3^4$ stand out, in the sense they are the ones for which any ring cuts will always lead to the same result. Curiously, these correspond exactly to the two inequivalent classes from the SLOCC classification in \cite{Dur:2000}, represented by the GHZ and W states, respectively.

\section{Application to four qubits}
\label{4q}

We shall again be concerned with the case where all initial qubits are entangled, i.e. with four-partite entanglement, for which our approach leads to 40 classes. Finding all entanglement classes is equivalent to finding all the polynomials $\ma{P}$ for four rings. As in Sec.~\ref{3q}, this can be done by direct inspection or computationally. All distinct polynomials and 1-cut diagrams are listed in Appendix \ref{appA}, as well as the corresponding states obtained using the procedure of Sec.~\ref{qumap}. It can be seen that the state coefficients are usually $1$, except when the polynomial contains a W type submonomial of Eq.~(\ref{W}) in it, in which case some coefficients of $1/2$ need to be introduced.

Amidst the multitude of classes obtained, a number of them stand out. For example, the Brunnian configuration $4^1$ and the generalized W type class $4^{40}$. These also belong to a group of six links that have the property of resulting in the same link when any ring is cut, the remaining ones being $4^6$, $4^{10}$, $4^{34}$ and $4^{38}$; one for each of the possible results $0$, $2^1$, $3^1$ $3^2$, $3^3$, and $3^4$, that occur after cutting any one ring from a 4-ring link. Consequently, by the same reasoning, there will be 46 classes with the same property when links with five rings are considered. Despite these type of classes having matched the SLOCC classes for three qubits, there is currently no evidence that there is some relation with SLOCC classification for the four qubit case.

Some clarification should be made for the classes $4^{22}$ and $4^{25}$. Although they have the same one cut diagrams, they are not equivalent. This can be checked immediately by noting that the two polynomials cannot be transformed into one another by a simple relabeling of variables, so the two classes are not equivalent. In order to distinguish between both of them, when tracing out the corresponding states, one only needs to observe that the possible results from the cuts are $cd$, $abd + ab$, and $abc + ab$ for $\ma{P}(4^{22})$, while for $\ma{P}(4^{25})$ they are $ac$, $bd$, $abc + ac$, and $abd + cd$.

Finally, it is essential to highlight the difficulties inherent to the detection of entanglement after each qubit trace, namely the possibility of bound entanglement when testing separability of mixed densities. We shall explore all the nuances by explicitly checking the correspondence of the state of Eq.~(\ref{420ket}), i.e.
\eq{
\rrho_{abcd}(4^{20}) = {\textrm{tr}_e \left[\ket{\psi_{20}}\bra{\psi_{20}}\right] \over \sqrt{\braket{\psi_{20} | \psi_{20}}}}\,,  \nonumber
}
where
\eq{
\ket{\psi_{20}} = \ket{3^1}_{abc}\ket{0}_d\ket{0}_e + \ket{3^1}_{abd}\ket{0}_c\ket{1}_e + \ket{2^1}_{ac}\ket{1 0}_{bd}\ket{2}_e \nonumber\,,
}
with the link class $4^{20}$, described by the polynomial ${abc + abd + ac}$, given in Eq.~(\ref{420pol}). As outlined in Sec.~\ref{qutolink}, we must first check entanglement of all initial qubits, then perform all possible traces and finally check which subsystems in the density operators after each trace are entangled or separable.

To check four-partite entanglement we begin by partial transposing the qubit $a$ in the density operator, obtaining the operator $\rrho^{T_a}_{abcd}$. We refrain from explicitly writing the operator for simplicity. The eigenvalues of $\rrho^{T_a}_{abcd}$ contain negative values, so in this case the PPT test guarantees that the qubit $a$ is entangled with the subsystem made of the remaining qubits. Repeating the same test for the operators $\rrho^{T_{b}}_{abcd}$, $\rrho^{T_c}_{abcd}$, $\rrho^{T_d}_{abcd}$, and $\rrho^{T_{ab}}_{abcd}$ reveals that all possible subsystems are entangled, so we check that the initial state is four-partite entangled, which is equivalent to considering the term $abcd$ in the polynomial.

Now that we have verified four-partite entanglement in the initial state, we must perform all possible traces and check if the results are compatible with the 1-cut diagram of Fig.~\ref{allcuts}. Tracing out the qubit $a$ first, we obtain the operator
\ea{\label{bcd}
\rrho_{bcd} = & {1\over 2}(2\ket{000}\bra{000}+2\ket{110}\bra{110}+\ket{100}\bra{100}+ \nonumber \\
              & \hspace{5mm} + \ket{101}\bra{101} )\,,
}
where, for simplificty, we refrain from labeling the qubits, which, whenever omitted, are assumed to be in alphabetic order. In this case, since the operator (\ref{bcd}) is diagonal and is written in the computational basis, we know that it is separable. However, let us suppose it was not diagonal, so separability was not clear. To find which subsystems were entangled or not, we would have to calculate the eigenvalues of the operators $\rrho^{T_b}_{bcd}$, $\rrho^{T_c}_{bcd}$ and $\rrho^{T_d}_{bcd}$. After doing so, we would be faced with only non-negative values for all cases, which would be an indication that all subsystems were separable.  However, there would be no guarantee of this since we would be dealing with $3 \times 3$ systems. In other words, the subsystems could still be entangled even though the states were PPT. What we could then do would be to calculate the eigenvectors of $\rrho_{bcd}$ and check if all of them corresponded to separable vectors, in which case it would be guaranteed that the density operator was separable, and so would all of its subsystems. Had it been the case that is was not immediately obvious the eigenstates were separable, we would have simply dismissed the initial state altogether and tried a different one. The states of Appendix~\ref{appA} are all optimized in this regard. In terms of links, the information that $\rrho_{bcd}$ is separable means that, once the ring $a$ is cut, none of the rings $b$, $c$ and $d$ are linked, i.e. the terms $bcd$, $bc$, $bd$ and $bd$ are not present in the polynomial.

Tracing now the qubit $b$ from the initial state, we obtain the operator
\ea{\label{acd}
\rrho_{acd} = & {1\over 2}\bigg(\left(\ket{000}+\ket{110}\right)\left(\bra{000}+\bra{110}\right)+2\ket{000}\bra{000}+ \nonumber \\
              & \hspace{5mm} + \ket{101}\bra{101}+\ket{110}\bra{110} \bigg)\,.
}
The eigenvalues of $\rrho^{T_a}_{acd}$ and $\rrho^{T_c}_{acd}$ contain negative values, while the eigenvalues of $\rrho^{T_d}_{acd}$ are non-negative. Consequently, the subsystems $a-cd$ and $c-ad$ are entangled, while $ac-d$ might be separable. To be certain, we compute the eigenstates $\ket{v_i}_{acd}$ of the density operator $\rrho_{acd}$ and build the associated pure density operators ${\rrho^{i}_{acd}=\ket{v_i}\bra{v_i}}$.
If, by tracing out qubit $d$ in each of them, all the reduced density operators remain pure, then each eigenvector can be decomposed as $\ket{v_i}_{acd} = \ket{v'_i}_{ac} \otimes \ket{v''_i}_{d}$. This is indeed the situation in this case, so we are able to write
\eq{
\rrho_{acd} = \sum^{3}_{i=1} \lambda_i \ket{v'_i}_{ac} \bra{v'_i} \otimes \ket{v''_i}_{d} \bra{v''_i}\,,
}
where $\lambda_i$ is the eigenvalue associated to the eigenvector $\ket{v_i}_{acd}$. In other words, the subsystem $d$ is separable from the subsystem $d-ac$. This guarantees that there is no tripartite entanglement, i.e. one of the rings is detached after the ring $b$ is cut from the original 4-ring link, implying that the terms $acd$, $ad$, and $cd$ are automatically ruled out from the polynomial. The only possible term which is not yet ruled out is $ac$. The presence of the term $ac$ is checked by a PPT test on $\rrho_{ac}$.

Regarding the trace of qubit $c$ from the initial state, we obtain the density operator
\ea{\label{abd}
\rrho_{abd} = & {1\over 2}\bigg(\left(\ket{000}+\ket{111}\right)\left(\bra{000}+\bra{111}\right)+ \nonumber \\
& \hspace{5mm} + \ket{000}\bra{000}+\ket{010}\bra{010}+2\ket{110}\bra{110} \bigg)~\,.
}
All of the partially transposed operators $\rrho^{T_{a}}_{abd}$, $\rrho^{T_{b}}_{abd}$ and $\rrho^{T_{d}}_{abd}$ have at least one negative eigenvalue, so we have tripartite entanglement, which means the term $abd$ must be included. Tracing further qubits will always result in separable operators, as confirmed by PPT tests, proving that the terms $ab$, $ad$ and $bd$ must not appear.

\begin{figure}[h!]
  \centering
    \includegraphics[width=0.27\textwidth]{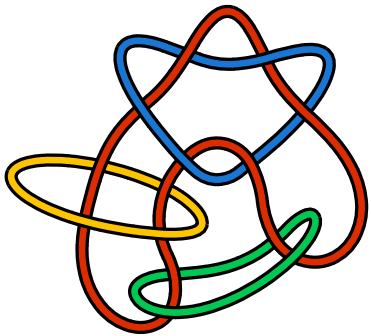}
      \caption{An element of the link class $4^{20}$.}
      \label{420r}
\end{figure}

Finally, tracing the qubit $d$ results in
\ea{\label{abc}
\rrho_{abc} = & {1\over 2}\bigg(\left(\ket{000}+\ket{111}\right)\left(\bra{000}+\bra{111}\right)+\ket{000}\bra{000}+\nonumber \\
& \hspace{1mm} + \left(\ket{010}+\ket{111}\right)\left(\bra{010}+\bra{111}\right)+\ket{110}\bra{110} \bigg)\,.
}
The partially transposed operators $\rrho^{T_a}_{abd}$, $\rrho^{T_b}_{abc}$ and $\rrho^{T}_{abc}$ all have non-negative eigenvalues, indicating that the term $abc$ is to be included. The remaining traces result in the operators $\rrho_{ab}$, $\rrho_{ac}$ and $\rrho_{bc}$, where only $\rrho_{ac}$ is entangled, revealing that only the term $ac$ must be included, as we already knew from the traces of Eq.~(\ref{acd}). In the end, we find the polynomial must be $abcd + abc + abd + ac$, although the $5^{\rm th}$ rule of polynomial construction of Sec.~\ref{form} dictates the term $abcd$ is superfluous, as all of its variables are already present in the lesser order monomial $abc + abd + ac$, associated to a 4-ring link class. We are left with the final polynomial $abc + abd + ac$, as required. The associated link class $4^{20}$ is represented in Fig.~\ref{420r}, where its properties under ring cuts can be straightforwardly visualized.


\section{Link representation}{}
\label{linkrep}

\begin{figure}[h!]
  \centering
    \includegraphics[width=0.40\textwidth]{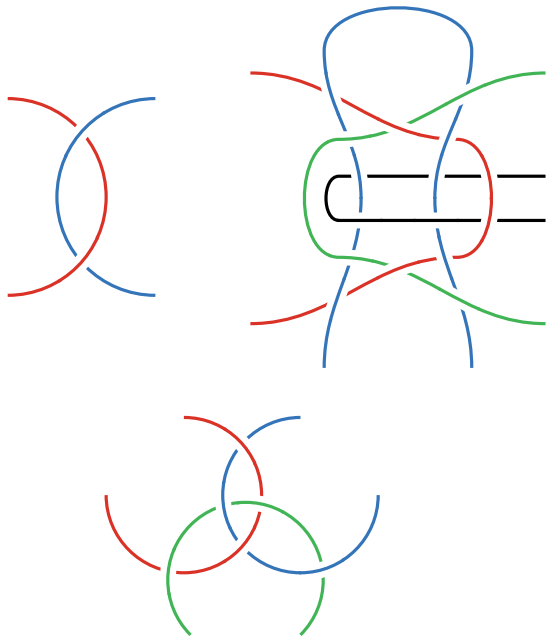}
      \caption{Basic building blocks of 4-ring links: on the top left the 2-variable term block; on the top right the 4-variable term block; and on the bottom the 3-variable term block.}
      \label{AllBlocks}
\end{figure}

In this section we will outline a way to represent a link using its polynomial representation. Although this is not especially important from the physical point of view, it is of relevance from the perspective of knot theory. The procedure uses basic building blocks associated to each term, changing according to the number of variables present. These are represented in Fig.~\ref{AllBlocks}, up to four variables, with the style used in knot theory. For any number of rings, a block is obtained by representing a Brunnian link, pulling out and cutting each ring, thus leaving two ends to be connected.

For any given polynomial, we start by choosing a color for each ring, draw each block independently for each term of the polynomial and finish by connecting all the lines of each respective color, without crossing any of them. Inserting the resulting knot in a program such as KnotPlot will simplify much further the visualization. This was the method used to generate all the links depicted in this paper.

\section{Physical applications for qubit networks}{}
\label{physapp}

Classification of quantum entanglement is usually directed to protocol applications. The goal of classification by SLOCC, for example, is to identify equivalence classes between quantum states which can in principle be used for the same protocols. The classification scheme developed in this paper, on the other hand, is instead concerned with whether the subsystems of a state may or may not perform protocols successfully with each other. This naturally suggests applications related to qubit networks, for which we will highlight some important aspects in this section.

In order to better understand where this new classification scheme distinguishes itself from others, one should look at situations where the partial trace of subsystems plays an important role. Consider, for instance, three qubits labeled $a$, $b$ and $c$, belonging to Alice, Bob and Charlie, respectively. These qubits belong to a tripartite entangled state, described by the density operator $\rrho_{abc}$. Now imagine that Charlie does not want to act on his qubit at all, while Alice and Bob decide to execute some protocol, completely ignoring Charlie's existence. The outcome of the protocol will be independent of whether Alice and Bob have full knowledge of the three qubit state or use instead the state $\rrho_{ab} = \textrm{tr}_c[{\rrho_{abc}}]$ where Charlie's qubit has been traced out. Indeed, in this situation, a general operation in the system will be of the form
\eq{
\hat{\ma{O}}(\rrho_{abc}) = \sum_{i} (\hat{M}_{i,ab} \otimes \hat{1}_{c}) \rrho_{abc} (\hat{M}^{\dagger}_{i,ab} \otimes \hat{1}_{c})
}
where the operators $\hat{M}_{i,ab}$ act on the Hilbert space of the subsystem Alice-Bob. Any protocol is given by a specific choice for the $\hat{M}_{i,ab}$ acting on the system, so the probability $p_i$ of obtaining the outcome $i$ from some protocol will be
\ea{
p_i & = \textrm{tr}_{abc}\left[ (\hat{M}^{\dagger}_{i,ab} \otimes \hat{1}_{c}) (\hat{M}_{i,ab} \otimes \hat{1}_{c}) \rrho_{abc}\right] \nonumber \\
& = \textrm{tr}_{ab}\left[ \hat{M}^{\dagger}_{i,ab} \hat{M}_{i,ab} \rrho_{ab}\right]\,.
}
Thus, in a situation where Charlie is inactive, the outcomes of protocols that Alice and Bob execute will not be affected by whether or not they had knowledge of the full tripartite state. Thus, as long as the state $\rrho_{ab}$ is entangled, Alice and Bob will be able to successfully carry out protocols. If Charlie decides to act, then the general operator acting on the system will be constructed from operators of the form $(\hat{M}_{i,ab} \otimes \hat{N}_{i,c})$. In this case, in order for Alice and Bob to successfully apply a protocol, they will have to know the operation $\hat{N}_{i,c}$ that Charlie decided to apply on his qubit.

An important consequence of the previous remarks is that Alice and Bob will not be able to perform protocols successfully  if $\rrho_{ab}$ is separable, with Charlie inactive, since no entanglement effectively exists between Alice's and Bob's qubits in that case. For all purposes, even if Charlie chooses to act on his qubit, if he does not transmit any information to anyone, Alice and Bob will not be able to identify correlations between their outcomes, thereby rendering any protocols between them useless and effectively cutting off their entanglement based communications. This is an important aspect to take into account when constructing qubit networks, since the entity responsible for building the network may want to completely incapacitate some parts from communicating with each other using their qubits, without the intervention of someone else. It is also a situation where the classification scheme developed in this work is specially suited for, since it revolves around the properties of quantum states under partial trace of subsystems.

To put the idea into practice, consider the following example. Imagine we want to build a qubit network of three qubits, distributed among Alice, Bob and Charlie, with the restriction that Bob and Charlie may never successfully perform any protocols without external help from Alice, while Alice may have a chance to communicate with either Bob or Charlie if she wishes to. In terms of density operators, the problem is to find a three-qubit state such that $\rrho_{abc}$ is entangled, the reduced density matrices $\rrho_{ab}$ and $\rrho_{ac}$ are entangled, and $\rrho_{bc}$ is separable. From the results of the previous sections, this implies that we are searching for a link class described by the polynomial ${abc + ab + ac}$, which is equivalent to ${ab + ac}$. This polynomial represents the link class $3^3$, so we immediately know the state of Eq.~(\ref{33ket}) will contain the properties we are interested in. The main advantage of the classification scheme using links is that the previous described task of finding the right 3-qubit state could have been performed using only ring variables from the start. Indeed, the task is analogous to finding a link class such that the ring $a$ is connected to $b$ and $c$ separately, implying the presence of the terms $ab$ and $ac$ in the polynomial, while the rings $b$ and $c$ are not linked directly, i.e. they don't remain linked after the ring $a$ is cut. Once we conclude the polynomial we need is ${ab + ac}$, we simply use the map between polynomials and quantum states to obtain an example of a suitable state.

The classification of entanglement using links is more relevant as the number of qubits increase. The advantage is not to determine the minimum amount of classes necessary for protocols, but rather to manage communications within qubit networks, i.e. to decide which parties must not be able to communicate with each other without external help. Already for four qubits the situation becomes non-trivial if entanglement classification through links is not used. For instance, consider four qubits distributed among Alice, Bob, Charlie and Diana, and imagine we want to build a network where the only communications between parties which are not completely cut off (without external action) are the following:
\begin{itemize}
\item Alice, Bob and Charlie;
\item Alice, Bob and Diana;
\item Alice and Charlie.
\end{itemize}
In terms of connections between links, the first restriction implies we need the term $abc$, the second restriction indicates the presence of the term $abd$ and the last restriction points out the presence of the term $ac$. Therefore, the polynomial we need is $abc + abd + ac$, which represents the class $4^{20}$. A representative state of this class is given in Eq.~(\ref{420ket}), which has already been verified to have the required properties under partial traces, in Sec~\ref{4q}. By using this state, we make sure that communications solely between Bob and Diana, for example, will never be possible, because the state is built in a way such that the reduced density operator $\rrho_{bd}$ is separable. If we were to address this problem without the aid of link classification, it would be much more challenging.

From the point of view of physical applications, it would be optimal if one had access to a map between polynomials and pure states, instead of mixed states. However, the fact we can quickly obtain mixed state representatives for each link class is already a significant step towards that goal.

\section{Conclusions}{}
\label{conc}

In this work we have developed a consistent formalism to classify quantum entanglement by associating particles with non-rigid rings and quantum entanglement with linked rings. Focusing on whether or not a ring is linked to another ring after a cut, rather than the linking intricacies, we were able to establish a simple equivalence criterion between links. The bridge with quantum mechanics was then established by considering the act of cutting a ring to be analogous to the trace of a particle from a state.

The idea was formalized by a set a rules which allowed the construction of a polynomial for each different link and a systematic procedure to associate links to quantum states and qubit states to links. We applied the formalism to three qubits, uncovering a new type of link class not mentioned until now, and to 4 qubits, recovering a vast amount of new classes. We have also described a brief procedure which facilitates the visualization of links by using their polynomial representation. The end result is a classification scheme with an intuitive picture of the entanglement classes, equipped with a set of procedures which quickly allow the identification of an example state for any class. We also demonstrated the physical potential of this new classification scheme by applying it to a management problem of communication between parties inside a qubit network.

A number of interesting future developments are in order, such as finding a map from link classes to pure quantum states or understanding more physical setups where entanglement classification through links plays a prominent role. One can also focus on developing increasingly efficient ways to find all polynomials for a given number of rings. However, this is not particularly relevant since the number of classes increases very rapidly with the number of rings, and the important aspect is to find which polynomial is associated to a given link. For example, for five rings one already finds exactly $6900$ different classes \footnote{A list of all polynomials associated to 5-ring links is available at https://fenix.tecnico.ulisboa.pt/homepage/ist165680.}. On more speculative grounds, one may inquire whether the details of the links, such as crossing numbers, may induce class subdivisions, or if other important quantities like entanglement entropy have a topological picture as well.


\appendix
\section{}{}
\label{appA}

For 4-ring links, we find the following classes:

\ea{
\ma{P}({4^1})    & = abcd\,, \\
\ma{P}({4^2})    & = abcd + abc\,, \\
\ma{P}({4^3})    & = abcd + abc + ab\,, \\
\ma{P}({4^4})    & = abcd + ab\,, \\
\ma{P}({4^5})    & = abcd + ab + ac\,, \\
\ma{P}({4^6})    & = abcd + ab + cd\,, \\
\ma{P}({4^7})    & = abcd + ab + ac + bc\,, \\
\ma{P}({4^8})    & = abc + abd\,, \\
\ma{P}({4^9})    & = abc + abd + acd\,, \\
\ma{P}({4^{10}}) & = abc + abd + acd + bcd\,, \\
\ma{P}({4^{11}}) & = abc + ad\,, \\
\ma{P}({4^{12}}) & = abc + ab + ad\,, \\
\ma{P}({4^{13}}) & = abc + ad + bd\,, \\
\ma{P}({4^{14}}) & = abc + ab + cd\,, \\
\ma{P}({4^{15}}) & = abc + ad + bd + cd\,, \\
\ma{P}({4^{16}}) & = abc + ab + ad + bd \,, \\
\ma{P}({4^{17}}) & = abc + ab + ad + cd\,, \\
\ma{P}({4^{18}}) & = abc + ab + ad + bd + cd\,, \\
\ma{P}({4^{19}}) & = abc + abd + ab\,, \\
\ma{P}({4^{20}}) & = abc + abd + ac\,, \label{420pol} \\
\ma{P}({4^{21}}) & = abc + abd + cd\,, \\
\ma{P}({4^{22}}) & = abc + abd + ab + cd\,, \\
\ma{P}({4^{23}}) & = abc + abd + ac + ad\,, \\
\ma{P}({4^{24}}) & = abc + abd + ac + cd\,, \\
\ma{P}({4^{25}}) & = abc + abd + ac + bd\,,
}
\ea{
\ma{P}({4^{26}}) & = abc + abd + ac + ad + cd\,, \\
\ma{P}({4^{27}}) & = abc + abd + ac + bd + cd\,, \\
\ma{P}({4^{28}}) & = abc + abd + acd + ab\,, \\
\ma{P}({4^{29}}) & = abc + abd + acd + bc\,, \\
\ma{P}({4^{30}}) & = abc + abd + acd + ab + cd\,, \\
\ma{P}({4^{31}}) & = abc + abd + acd + bc + bd\,, \\
\ma{P}({4^{32}}) & = abc + abd + acd + bc + bd + cd\,, \\
\ma{P}({4^{33}}) & = abc + abd + acd + bcd + ab\,, \\
\ma{P}({4^{34}}) & = abc + abd + acd + bcd + ab + cd\,, \\
\ma{P}({4^{35}}) & = ab + ac + ad\,, \\
\ma{P}({4^{36}}) & = ab + ac + bd\,, \\
\ma{P}({4^{37}}) & = ab + ac + ad + bc\,, \\
\ma{P}({4^{38}}) & = ab + ac + bd + cd\, \\
\ma{P}({4^{39}}) & = ab + ac + ad + bc + bd\,, \\
\ma{P}({4^{40}}) & = ab + ac + ad + bc + bd + cd\,.
}

A mixed state associated to each of the previous polynomials is defined by substituting $\ket{\psi_i}$ in
\eq{\label{rhopsi4}
\rrho_{abcd}(4^i) = {\textrm{tr}_e \left[\ket{\psi_i}\bra{\psi_i}\right] \over \sqrt{\braket{\psi_i | \psi_i}}}\,,
}
the following states:
\ea{
\ket{\psi_1}    = & \ket{4^1}_{abcd}\ket{0}_e\,, \\
\ket{\psi_2}    = & \ket{4^1}_{abcd}\ket{0}_e + \ket{3^1}_{abc}\ket{0}_d\ket{1}_e\,, \\
\ket{\psi_3}    = & \ket{4^1}_{abcd}\ket{0}_e + \ket{3^1}_{abc}\ket{1}_d\ket{1}_e + \nonumber \\
               & + \ket{2^1}_{ab}\ket{0 0}_{cd}\ket{2}_e\,, \\
\ket{\psi_4}    = & \ket{4^1}_{abcd}\ket{0}_e + \ket{2^1}_{ab}\ket{0 1}_{cd}\ket{1}_e\,, \\
\ket{\psi_5}    = & \ket{4^1}_{abcd}\ket{0}_e + \ket{2^1}_{ab}\ket{0 0}_{cd}\ket{1}_e + \nonumber \\
               & + \ket{2^1}_{ac}\ket{0 1}_{bd}\ket{2}_e\,, \\
\ket{\psi_6}    = & \ket{4^1}_{abcd}\ket{0}_e + \ket{2^1}_{ab}\ket{0 0}_{cd}\ket{1}_e + \nonumber \\
               & + \ket{2^1}_{cd}\ket{0 1}_{ab}\ket{2}_e\,, \\
\ket{\psi_7}    = & {1\over 2}\ket{4^1}_{abcd}\ket{0}_e + \ket{2^1}_{ab}\ket{1 0}_{cd}\ket{1}_e + \nonumber \\
               & + {1\over 2} \ket{2^1}_{ac}\ket{0 1}_{bd}\ket{2}_e + {1\over 2} \ket{2^1}_{bc}\ket{0 1}_{ad}\ket{3}_e\,, \\
\ket{\psi_8}    = & \ket{3^1}_{abc}\ket{1}_d\ket{0}_e + \ket{3^1}_{abd}\ket{0}_c\ket{1}_e\,, \\
\ket{\psi_9}    = & \ket{3^1}_{abc}\ket{0}_d\ket{0}_e + \ket{3^1}_{abd}\ket{0}_c\ket{1}_e + \nonumber \\
               & + \ket{3^1}_{acd}\ket{0}_b\ket{2}_e\,, \\
\ket{\psi_{10}} = & \ket{3^1}_{abc}\ket{0}_d\ket{0}_e + \ket{3^1}_{abd}\ket{0}_c\ket{1}_e + \nonumber \\
               & + \ket{3^1}_{acd}\ket{0}_b\ket{2}_e + \ket{3^1}_{bcd}\ket{0}_a\ket{3}_e\,, \\
\ket{\psi_{11}} = & \ket{3^1}_{abc}\ket{0}_d\ket{0}_e + \ket{2^1}_{ad}\ket{0 0}_{bc}\ket{1}_e\,,
               }
\ea{
\ket{\psi_{12}} = & \ket{3^1}_{abc}\ket{0}_d\ket{0}_e + \ket{2^1}_{ad}\ket{0 0}_{bc}\ket{1}_e + \nonumber \\
               & + \ket{2^1}_{ab}\ket{0 0}_{cd}\ket{2}_e\,, \\
\ket{\psi_{13}} = & \ket{3^1}_{abc}\ket{0}_d\ket{0}_e + \ket{2^1}_{ad}\ket{1 1}_{bc}\ket{1}_e + \nonumber \\
               & + \ket{2^1}_{bd}\ket{1 0}_{ac}\ket{2}_e\,, \\
\ket{\psi_{14}} = & \ket{3^1}_{abc}\ket{0}_d\ket{0}_e + \ket{2^1}_{ab}\ket{0 0}_{cd}\ket{1}_e + \nonumber \\
               & + \ket{2^1}_{cd}\ket{1 0}_{ab}\ket{2}_e\,, \\
\ket{\psi_{15}} = & \ket{3^1}_{abc}\ket{0}_d\ket{0}_e + \ket{2^1}_{ad}\ket{1 1}_{bc}\ket{1}_e + \nonumber \\
               & + \ket{2^1}_{bd}\ket{1 1}_{ac}\ket{2}_e + \ket{2^1}_{cd}\ket{1 1}_{ab}\ket{3}_e\,, \\
\ket{\psi_{16}} = & {1 \over 2}\ket{3^1}_{abc}\ket{1}_d\ket{0}_e + \ket{2^1}_{ab}\ket{1 1}_{cd}\ket{1}_e + \nonumber \\
               & + \ket{2^1}_{ad}\ket{0 1}_{bc}\ket{2}_e + {1 \over 2} \ket{2^1}_{bd}\ket{0 0}_{ac}\ket{3}_e\,, \\
\ket{\psi_{17}} = & \ket{3^1}_{abc}\ket{1}_d\ket{0}_e + \ket{2^1}_{ab}\ket{0 1}_{cd}\ket{1}_e + \nonumber \\
               & + \ket{2^1}_{ad}\ket{1 0}_{bc}\ket{2}_e + \ket{2^1}_{cd}\ket{0 1}_{ab}\ket{3}_e\,, \\
\ket{\psi_{18}} = & {1 \over 2}\ket{3^1}_{abc}\ket{0}_d\ket{0}_e + {1 \over 2}\ket{2^1}_{ab}\ket{0 0}_{cd}\ket{1}_e + \nonumber \\
               & + \ket{2^1}_{ad}\ket{1 1}_{bc}\ket{2}_e + {1 \over 2}\ket{2^1}_{bd}\ket{0 0}_{ac}\ket{3}_e + \nonumber \\
               & + \ket{2^1}_{cd}\ket{1 1}_{ab}\ket{4}_e \,, \\
\ket{\psi_{19}} = & \ket{3^1}_{abc}\ket{0}_d\ket{0}_e + \ket{3^1}_{abd}\ket{0}_c\ket{1}_e + \nonumber \\
               & + \ket{2^1}_{ab}\ket{1 0}_{cd}\ket{2}_e\,, \\
\ket{\psi_{20}} = & \ket{3^1}_{abc}\ket{0}_d\ket{0}_e + \ket{3^1}_{abd}\ket{0}_c\ket{1}_e + \nonumber \\
               & + \ket{2^1}_{ac}\ket{1 0}_{bd}\ket{2}_e\,, \label{420ket} \\
\ket{\psi_{21}} = & \ket{3^1}_{abc}\ket{1}_d\ket{0}_e + \ket{3^1}_{abd}\ket{0}_c\ket{1}_e + \nonumber \\
               & + \ket{2^1}_{cd}\ket{0 0}_{ab}\ket{2}_e\,, \\
\ket{\psi_{22}} = & \ket{3^1}_{abc}\ket{0}_d\ket{0}_e + \ket{3^1}_{abd}\ket{1}_c\ket{1}_e + \nonumber \\
               & + \ket{2^1}_{ab}\ket{1 1}_{cd}\ket{2}_e + \ket{2^1}_{cd}\ket{1 0}_{ab}\ket{3}_e\,, \\
\ket{\psi_{23}} = & \ket{3^1}_{abc}\ket{1}_d\ket{0}_e + \ket{3^1}_{abd}\ket{1}_c\ket{1}_e + \nonumber \\
               & + \ket{2^1}_{ac}\ket{1 1}_{bd}\ket{2}_e + \ket{2^1}_{ad}\ket{1 1}_{bc}\ket{3}_e\,, \\
\ket{\psi_{24}} = & \ket{3^1}_{abc}\ket{1}_d\ket{0}_e + \ket{3^1}_{abd}\ket{0}_c\ket{1}_e + \nonumber \\
               & + \ket{2^1}_{ac}\ket{1 1}_{bd}\ket{2}_e + \ket{2^1}_{cd}\ket{1 0}_{ab}\ket{3}_e\,, \\
\ket{\psi_{25}} = & \ket{3^1}_{abc}\ket{0}_d\ket{0}_e + \ket{3^1}_{abd}\ket{0}_c\ket{1}_e + \nonumber \\
               & + \ket{2^1}_{ac}\ket{1 1}_{bd}\ket{2}_e + \ket{2^1}_{bd}\ket{0 0}_{ac}\ket{3}_e\,, \\
\ket{\psi_{26}} = & {1\over 2} \ket{3^1}_{abc}\ket{0}_d\ket{0}_e + {1\over 2}\ket{3^1}_{abd}\ket{1}_c\ket{1}_e + \nonumber \\
               & + \ket{2^1}_{ac}\ket{0 0}_{bd}\ket{2}_e  + \ket{2^1}_{ad}\ket{0 1}_{bc}\ket{3}_e +\nonumber \\
               & + {1\over 2} \ket{2^1}_{cd}\ket{1 1}_{ab}\ket{4}_e\,, \\
\ket{\psi_{27}} = & \ket{3^1}_{abc}\ket{0}_d\ket{0}_e + \ket{3^1}_{abd}\ket{1}_c\ket{1}_e + \nonumber \\
               & + \ket{2^1}_{ac}\ket{1 0}_{bd}\ket{2}_e + \ket{2^1}_{bd}\ket{0 1}_{ac}\ket{4}_e + \nonumber \\
               & + \ket{2^1}_{cd}\ket{0 1}_{ab}\ket{3}_e \,, \\
\ket{\psi_{28}} = & \ket{3^1}_{abc}\ket{0}_d\ket{0}_e + \ket{3^1}_{abd}\ket{0}_c\ket{1}_e + \nonumber \\
               & + \ket{3^1}_{acd}\ket{0}_b\ket{2}_e + \ket{2^1}_{ab}\ket{0 0}_{cd}\ket{3}_e\,, \\
\ket{\psi_{29}} = & \ket{3^1}_{abc}\ket{1}_d\ket{0}_e + \ket{3^1}_{abd}\ket{1}_c\ket{1}_e + \nonumber \\
               & + \ket{3^1}_{acd}\ket{0}_b\ket{2}_e + \ket{2^1}_{bc}\ket{1 1}_{ad}\ket{3}_e\,,
                              }
\ea{
\ket{\psi_{30}} = & \ket{3^1}_{abc}\ket{1}_d\ket{0}_e + \ket{3^1}_{abd}\ket{0}_c\ket{1}_e + \nonumber \\
               & + \ket{3^1}_{acd}\ket{1}_b\ket{2}_e+\ket{2^1}_{ab}\ket{1 1}_{cd}\ket{3}_e + \nonumber \\
               & + \ket{2^1}_{cd}\ket{0 1}_{ab}\ket{4}_e\,, \\
\ket{\psi_{31}} = & \ket{3^1}_{abc}\ket{1}_d\ket{0}_e + \ket{3^1}_{abd}\ket{1}_c\ket{1}_e + \nonumber \\
               & + \ket{3^1}_{acd}\ket{0}_b\ket{2}_e + \ket{2^1}_{bc}\ket{1 1}_{ad}\ket{3}_e + \nonumber \\
               & + \ket{2^1}_{bd}\ket{0 1}_{ac}\ket{4}_e\,, \\
\ket{\psi_{32}} = & {1\over 2}\ket{3^1}_{abc}\ket{0}_d\ket{0}_e + {1\over 2}\ket{3^1}_{abd}\ket{1}_c\ket{1}_e + \nonumber \\
               & + {1\over 2}\ket{3^1}_{acd}\ket{0}_b\ket{2}_e + {1\over 2}\ket{2^1}_{bc}\ket{0 1}_{ad}\ket{3}_e + \nonumber \\
               & + \ket{2^1}_{bd}\ket{1 1}_{ac}\ket{4}_e + \ket{2^1}_{cd}\ket{0 0}_{ab}\ket{5}_e\,, \\
\ket{\psi_{33}} = & \ket{3^1}_{abc}\ket{0}_d\ket{0}_e + \ket{3^1}_{abd}\ket{0}_c\ket{1}_e + \nonumber \\
               & + \ket{3^1}_{acd}\ket{1}_b\ket{2}_e + \ket{3^1}_{bcd}\ket{0}_a\ket{3}_e + \nonumber \\
               & + \ket{2^1}_{ab}\ket{0 0}_{cd}\ket{4}_e\,,
                              }
\ea{
\ket{\psi_{34}} = & \ket{3^1}_{abc}\ket{1}_d\ket{0}_e + \ket{3^1}_{abd}\ket{0}_c\ket{1}_e + \nonumber \\
               & + \ket{3^1}_{acd}\ket{1}_b\ket{2}_e + \ket{3^1}_{bcd}\ket{0}_a\ket{3}_e + \nonumber \\
               & + \ket{2^1}_{ab}\ket{1 1}_{cd}\ket{4}_e + \ket{2^1}_{cd}\ket{1 1}_{ab}\ket{5}_e\,, \\
\ket{\psi_{35}} = & \ket{2^1}_{ab}\ket{1 1}_{cd}\ket{0}_e + \ket{2^1}_{ac}\ket{0 1}_{bd}\ket{1}_e + \nonumber \\
               & + \ket{2^1}_{ad}\ket{0 1}_{bc}\ket{2}_e\,, \\
\ket{\psi_{36}} = & \ket{2^1}_{ab}\ket{0 0}_{cd}\ket{0}_e + \ket{2^1}_{ac}\ket{1 1}_{bd}\ket{1}_e + \nonumber \\
               & + \ket{2^1}_{bd}\ket{0 0}_{ac}\ket{2}_e\,, \\
\ket{\psi_{37}} = & \ket{2^1}_{ab}\ket{0 0}_{cd}\ket{0}_e + {1\over 2}\ket{2^1}_{ac}\ket{0 0}_{bd}\ket{1}_e + \nonumber \\
               & + {1\over 2}\ket{2^1}_{ad}\ket{1 0}_{bc}\ket{2}_e + \ket{2^1}_{bc}\ket{1 1}_{ad}\ket{3}_e\,, \\
\ket{\psi_{38}} = & \ket{2^1}_{ab}\ket{1 1}_{cd}\ket{0}_e + \ket{2^1}_{ac}\ket{0 0}_{bd}\ket{1}_e + \nonumber \\
               & + \ket{2^1}_{bd}\ket{1 1}_{ac}\ket{2}_e + \ket{2^1}_{cd}\ket{0 0}_{ab}\ket{3}_e\,, \\
\ket{\psi_{39}} = & {1\over 2}\ket{2^1}_{ab}\ket{0 0}_{cd}\ket{0}_e + \ket{2^1}_{ac}\ket{0 0}_{bd}\ket{1}_e+\nonumber \\
               & + {1\over 2}\ket{2^1}_{ad}\ket{0 0}_{bc}\ket{2}_e + \ket{2^1}_{bc}\ket{1 1}_{ad}\ket{3}_e + \nonumber \\
               & + \ket{2^1}_{bd}\ket{1 1}_{ac}\ket{4}_e\,, \\
\ket{\psi_{40}} = & {1\over 2} \ket{2^1}_{ab}\ket{1 0}_{cd}\ket{0}_e + \ket{2^1}_{ac}\ket{0 0}_{bd}\ket{1}_e + \nonumber \\
               & + {1\over 2}\ket{2^1}_{ad}\ket{0 1}_{bc}\ket{2}_e + {1\over 2}\ket{2^1}_{bc}\ket{1 0}_{ad}\ket{3}_e + \nonumber \\
               & + \ket{2^1}_{bd}\ket{1 1}_{ac}\ket{4}_e + \ket{2^1}_{cd}\ket{1 0}_{ab}\ket{5}_e\,.
}

In Fig.~\ref{allcuts} we list all the possible 1-cut diagrams for each class of 4-ring links.

\vfill

\pagebreak

\begin{widetext}
$\,$
\begin{figure}[h!]
\includegraphics[width=0.9\textwidth]{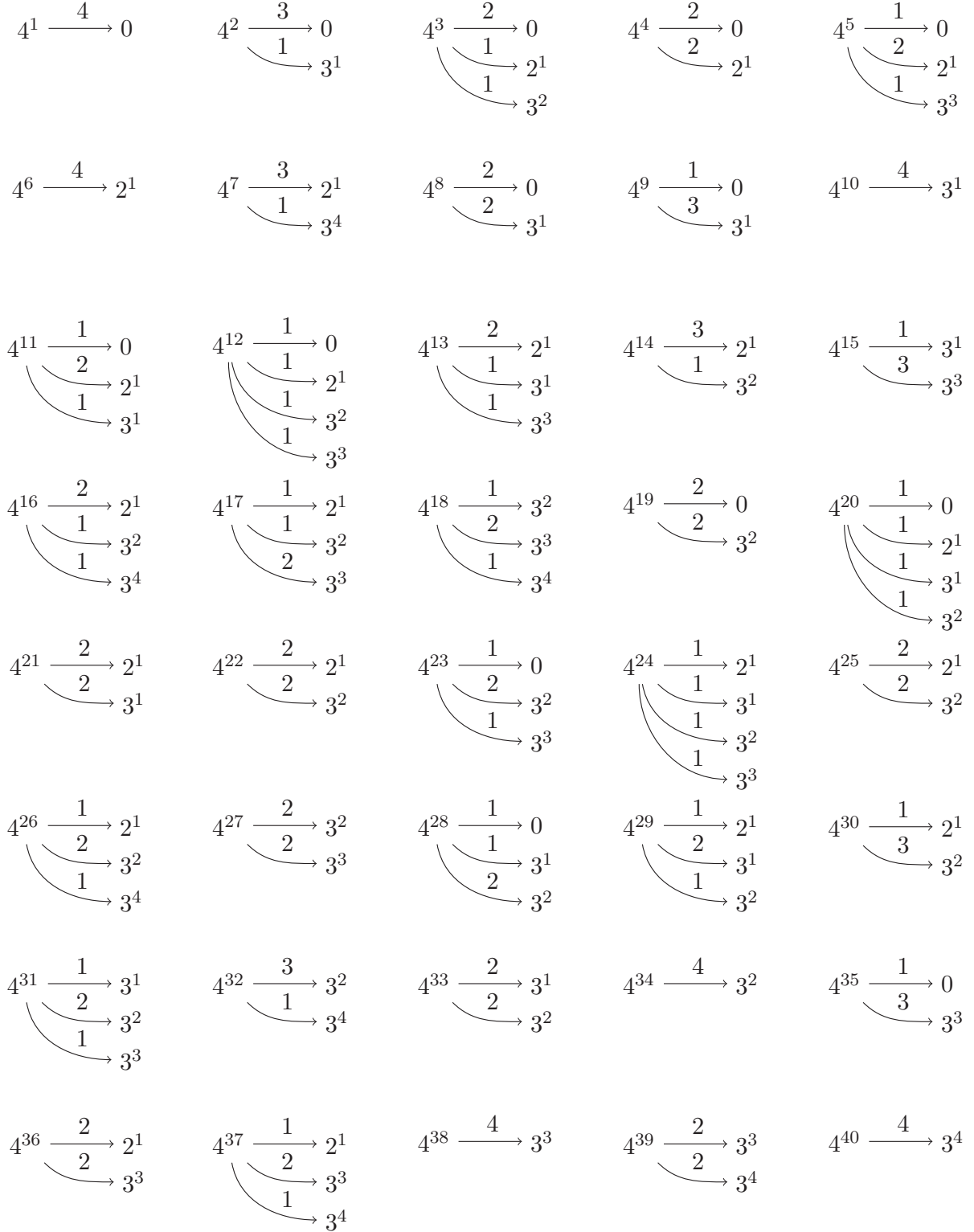}
\caption{All 1-cut diagrams corresponding to each 4-ring link class.}
\label{allcuts}
\end{figure}
$\,$
\end{widetext}

\vfill

\pagebreak

\acknowledgements
G.Q. acknowledges Grant No.~SFRH/BD/92583/2013 from Fundação para a Ciência e Tecnologia (FCT) - Portugal. R.A. acknowledges support from the Doctoral Programme in the Physics and Mathematics of Information (DP-PMI) and FCT through Scholarship No.~PD/BD/135011/2017.

G.Q. and R.A. also thank Nikola Paunkovi\'{c}, Mariano Lemus, Hadi Zahir and Jorge Lopes for helpful discussions.

\end{document}